\documentclass[aps,nofootinbib,preprintnumbers,amssymb,superscriptaddress,12pt]{revtex4}
\usepackage{amsmath}
\usepackage{graphicx}
\usepackage{epsfig}
\usepackage{multirow}
\usepackage{dcolumn}
\usepackage{bm}
\usepackage[usenames]{color}

\usepackage{epsfig}

\newcommand{\newc}{\newcommand}

\def\be{\begin{equation}}
\def\ee{\end{equation}}
\def\gs{\mathrel{
   \rlap{\raise 0.511ex \hbox{$>$}}{\lower 0.511ex \hbox{$\sim$}}}}
\def\ls{\mathrel{
   \rlap{\raise 0.511ex \hbox{$<$}}{\lower 0.511ex \hbox{$\sim$}}}}

\newcommand{\ba}{\begin{array}{c}}
\newcommand{\baz}{\begin{array}{cc}}
\newcommand{\barrr}{\begin{array}{rrr}}
\newcommand{\bad}{\begin{array}{ccc}}
\newcommand{\bav}{\begin{array}{cccc}}
\newcommand{\baf}{\begin{array}{ccccc}}
\newcommand{\bea}{\begin{equation} \begin{array}{c}}
\newcommand{\eea}{ \end{array} \end{equation}}
\newcommand{\ea}{\end{array}}
\newcommand{\D}{\displaystyle}

\newcommand{\gsim}{\raise0.3ex\hbox{$\;>$\kern-0.75em\raise-1.1ex\hbox{$\sim\;$}}}
\newcommand{\lsim}{\raise0.3ex\hbox{$\;<$\kern-0.75em\raise-1.1ex\hbox{$\sim\;$}}}
\newc{\nn}{\noindent}
\newc{\non}{\nonumber}

\def\beq{\begin{equation}}
\def\eeq{\end{equation}}
\def\baz{\begin{array}{cc}}
\def\ea{\end{array}}

\def\3e{e \bar{e} e}
\def\m2eg{\mu \to e \gamma}
\def\m2{\mu \to}
\def\m2e{\mu \to e}
\def\t2eg{\tau \to e \gamma}
\def\t2mg{\tau \to \mu \gamma}
\def\gs{\mathrel{
   \rlap{\raise 0.511ex \hbox{$>$}}{\lower 0.511ex \hbox{$\sim$}}}}
\def\ls{\mathrel{
   \rlap{\raise 0.511ex \hbox{$<$}}{\lower 0.511ex \hbox{$\sim$}}}}

\begin{document}
\mbox{ }\\[2cm]
\title{Simple two Parameter Description of Lepton Mixing}
\author{Werner Rodejohann}%
\email{werner.rodejohann@mpi-hd.mpg.de}
\author{He Zhang}%
\email{he.zhang@mpi-hd.mpg.de}
\affiliation{Max--Planck--Institut f\"ur Kernphysik,\\
Saupfercheckweg 1, 69117 Heidelberg, Germany}
%
\begin{abstract}
\noindent
We note that a simple two parameter description of lepton mixing is
possible which reproduces the features that apparently emerge from
global fits at the $1\sigma$ level: if $U_{e3}$ is non-zero it implies that
the solar neutrino mixing parameter $\sin^2 \theta_{12}$ is less than
$\frac 13$ by order $|U_{e3}|^2$. If the CP phase $\delta$ is around
$\pi$ it implies that the atmospheric neutrino mixing parameter
$\sin^2 \theta_{23}$ is less than $\frac 12$ by order $|U_{e3}|$.
The mixing scheme can be described by a 23-rotation appearing to the
right of a tri-bimaximal mixing matrix. We quantify the excellent
agreement of the scheme with data statistically, and comment on model building
aspects.

\end{abstract}


\maketitle

\newpage
\section{\label{sec:intro}Introduction}
\noindent All three mixing angles of the lepton sector are now
known. The last step towards this marvelous achievement came from
reactor neutrino experiments Double Chooz \cite{Abe:2011fz}, Daya
Bay \cite{An:2012eh} and RENO \cite{Ahn:2012nd}. Combining the
reactor data with other experiments ruled out vanishing $U_{e3}$ at
more than $7\sigma$ C.L.
\cite{Machado:2011ar,Tortola:2012te,Fogli:2012ua,thomas}. At the
Neutrino 2012 conference in June 2012, Double Chooz have presented
new data with $3.1\sigma$ evidence for non-zero $U_{e3}$, and also
Daya Bay has increased its significance to more than $7\sigma$, see
the URL {\tt
http://kds.kek.jp/conferenceTimeTable.py?confId=9151\#20120604.detailed}
for the slides. Moreover, additional data from T2K and MINOS can be
included, and one can fit the overall data to the parameters in the
Pontecorvo-Maki-Nakagawa-Sakata (PMNS) lepton mixing matrix
(ignoring possible Majorana phases)
\begin{eqnarray}\label{eq:para}
    U  = \left( \begin{matrix}c_{12} c_{13} & s_{12} c_{13} & s_{13} e^{-{\rm i}\delta} \cr -s_{12} c_{23}-c_{12} s_{23} s_{13} e^{{\rm i} \delta} & c_{12} c_{23}-s_{12} s_{23} s_{13} e^{{\rm i} \delta} & s_{23} c_{13} \cr s_{12} s_{23}-c_{12} c_{23} s_{13} e^{{\rm i} \delta} & -c_{12} s_{23}-s_{12} c_{23} s_{13} e^{{\rm i} \delta} & c_{23} c_{13} \end{matrix} \right) .
\end{eqnarray}
In this short note we wish to give a possible interpretation of the
emerging features of global fits. We will focus on the results from
Fogli {\it et al.} \cite{Fogli:2012ua}, which spotlight the following
interesting properties:
\begin{itemize}
\item[a)] $|U_{e3}| \simeq 0.16$ is sizable;
\item[b)] solar neutrino mixing is described by $\sin^2 \theta_{12}
\simeq \frac 13 - {\cal O}(|U_{e3}|^2)$, i.e.~slightly less than
$\frac 13$;
\item[c)] atmospheric neutrino mixing is described by $\sin^2 \theta_{23}
\simeq \frac 12 - {\cal O}(|U_{e3}|)$, i.e.~significantly less than
$\frac 12$;
\item[d)] the CP phase $\delta$ is around $\pi$.
\end{itemize}
The precise fit parameters are quoted in Table I. Not all features
are present in the results of the other groups
\cite{Machado:2011ar,Tortola:2012te,thomas} (which have not yet
updated their results, and have partly different treatment in their
atmospheric codes or do not even fit atmospheric data), and at the
$2\sigma$ level points b) and d) are absent. While there is no doubt
about the value of $|U_{e3}|$, less-than-maximal atmospheric mixing
and less-than-$\frac13$ solar mixing seem to be common features, at
least at the $1\sigma$ level.

If these properties of lepton mixing survive the
test of time, an interpretation in terms of a mixing scheme will
without doubt be useful. We note in this work that there is a mixing
scheme that can reproduce features a) -- d). It has only
two free parameters that can be adjusted to the observables, and possesses the following properties:
\begin{itemize}
\item[i)] if $|U_{e3}|$ is non-zero, $\sin^2 \theta_{12}
\simeq \frac 13 - {\cal O}(|U_{e3}|^2)$ is implied. This links
features a) and b);
\item[ii)] if $\delta$ is around $\pi$, $\sin^2 \theta_{23}
\simeq \frac 12 - {\cal O}(|U_{e3}|)$ is implied. This links features
c) and d).
\end{itemize}

The defining property for the PMNS matrix $U$ is
\be
|U| = \left(
\bad
\sqrt{\frac 23} & \# & \# \\
\sqrt{\frac 16} & \# & \# \\
\sqrt{\frac 16} & \# & \#
\ea
\right) ,
\ee
where the elements in the second and third column can be obtained by
unitarity. The entries of this matrix in the first column are exactly
as in tri-bimaximal mixing (TBM) \cite{tbm}, and hence this mixing matrix
can also be obtained by multiplying the tri-bimaximal mixing matrix
with a 23-rotation from the right. The phenomenology of the mixing scheme has been
analyzed first in \cite{Albright:2008rp} (see also \cite{He:2011gb}), and is a variant of the
so-called trimaximal mixing scheme, which is defined through a mixing
matrix with only the second row as for TBM
\cite{Bjorken:2005rm,He:2006qd,Lam:2006wm}.  This scheme is however disfavored
due to its prediction $\sin^2 \theta_{12} \simeq \frac 13 (1 +
|U_{e3}|^2)$. The scheme that we propose to describe the current data has been
written down first in \cite{Lam:2006wm}, and in the convention of
\cite{Albright:2008rp} is called TM$_1$. We revisit this mixing scheme
in this short note, emphasizing that it is able to perfectly
accommodate the above features a) -- d), which seem to emerge from global
fits. In Section \ref{sec:pheno} we study its phenomenology, and in
Section \ref{sec:theo} we discuss aspects of a possible theoretical
background.

\begin{table}[t]
\centering \caption{Best-fit and estimated 1$\sigma$ values of the
neutrino mixing parameters from Ref.~\cite{Fogli:2012ua} for the
normal mass ordering ($m_3> m_2 > m_1$, NH) and the inverted mass
ordering ($m_3<m_1<m_2$, IH).} \label{table:data} \vspace{8pt}
\begin{tabular}{c|cccc}
\hline \hline  Parameter & ~~~~~~$\sin^2\theta_{12}$ ~~~~~~& ~~~~~~$\sin^2\theta_{23}$ ~~~~~~ & ~~~~~~$\sin^2\theta_{13}$ ~~~~~~ & ~~~~~~$\delta/\pi$ ~~~~~~ \\
\hline NH & \multirow{2}{*}{$0.307^{+0.018}_{-0.016}$} & $0.386^{+0.024}_{-0.021}$ & $0.0241 \pm 0.0025 $ & $1.08^{+0.28}_{-0.31}$\\
IH & & $0.392^{+0.039}_{-0.022}$ & $0.0244^{+ 0.0023}_{-0.0025} $ & $1.09^{+0.38}_{-0.26}$ \\
\hline \hline
\end{tabular}
\end{table}

\section{\label{sec:pheno}Phenomenology}
\begin{center}
\begin{figure*}[hpt]
\includegraphics[width=10.035cm,height=6.950cm,angle=0]{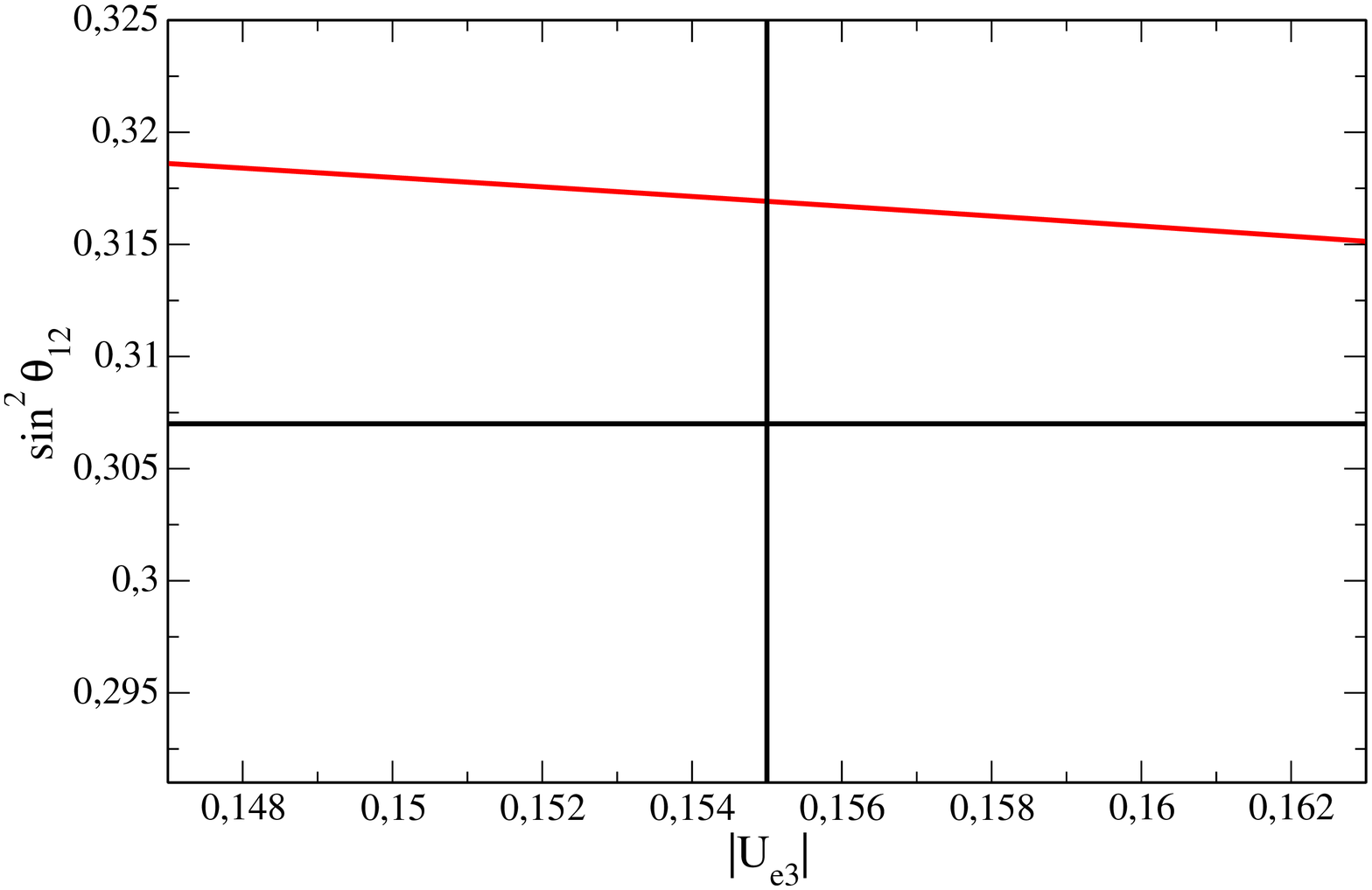}

\includegraphics[width=10.035cm,height=6.950cm,angle=0]{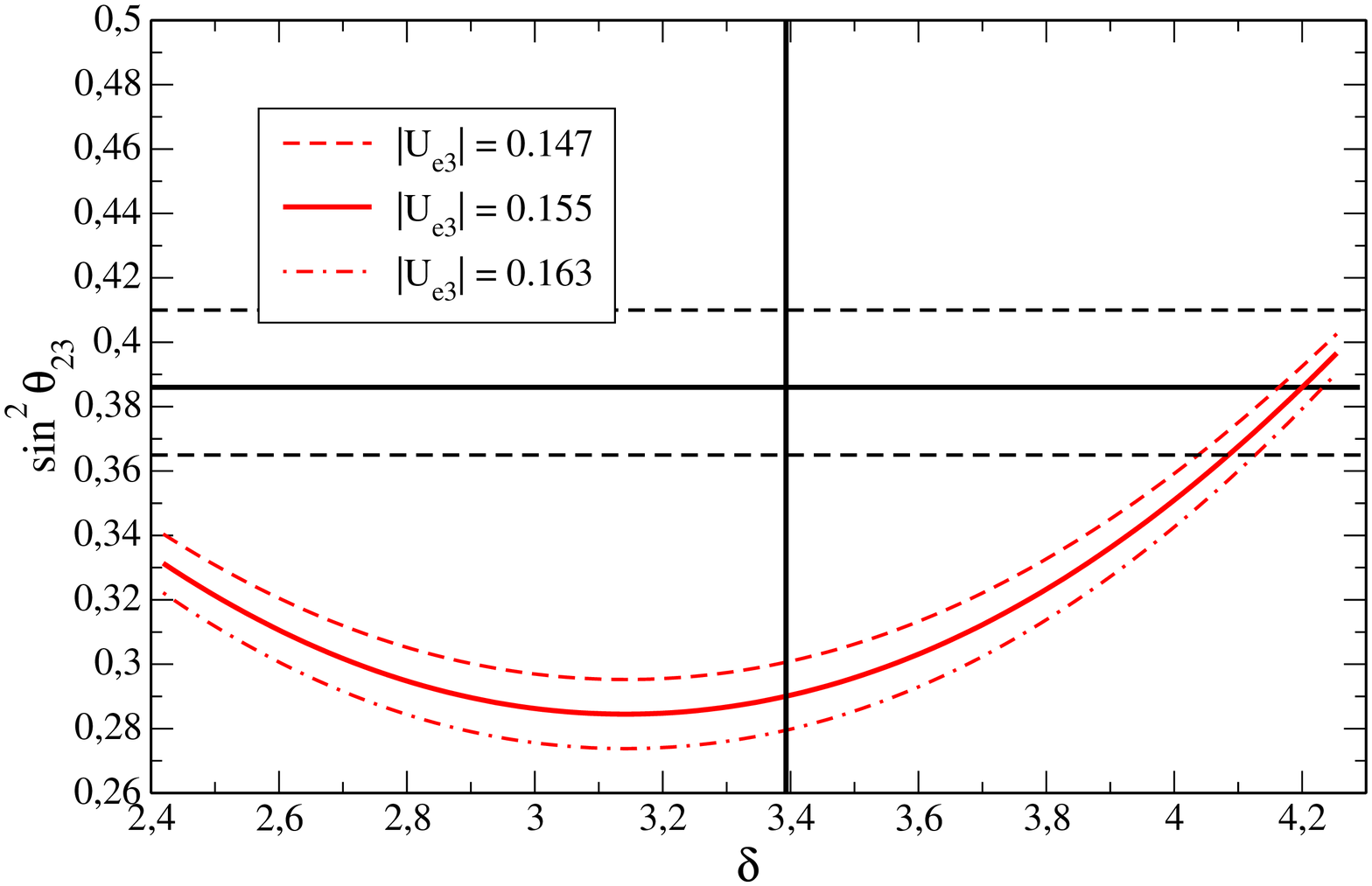}

\includegraphics[width=10.035cm,height=6.950cm,angle=0]{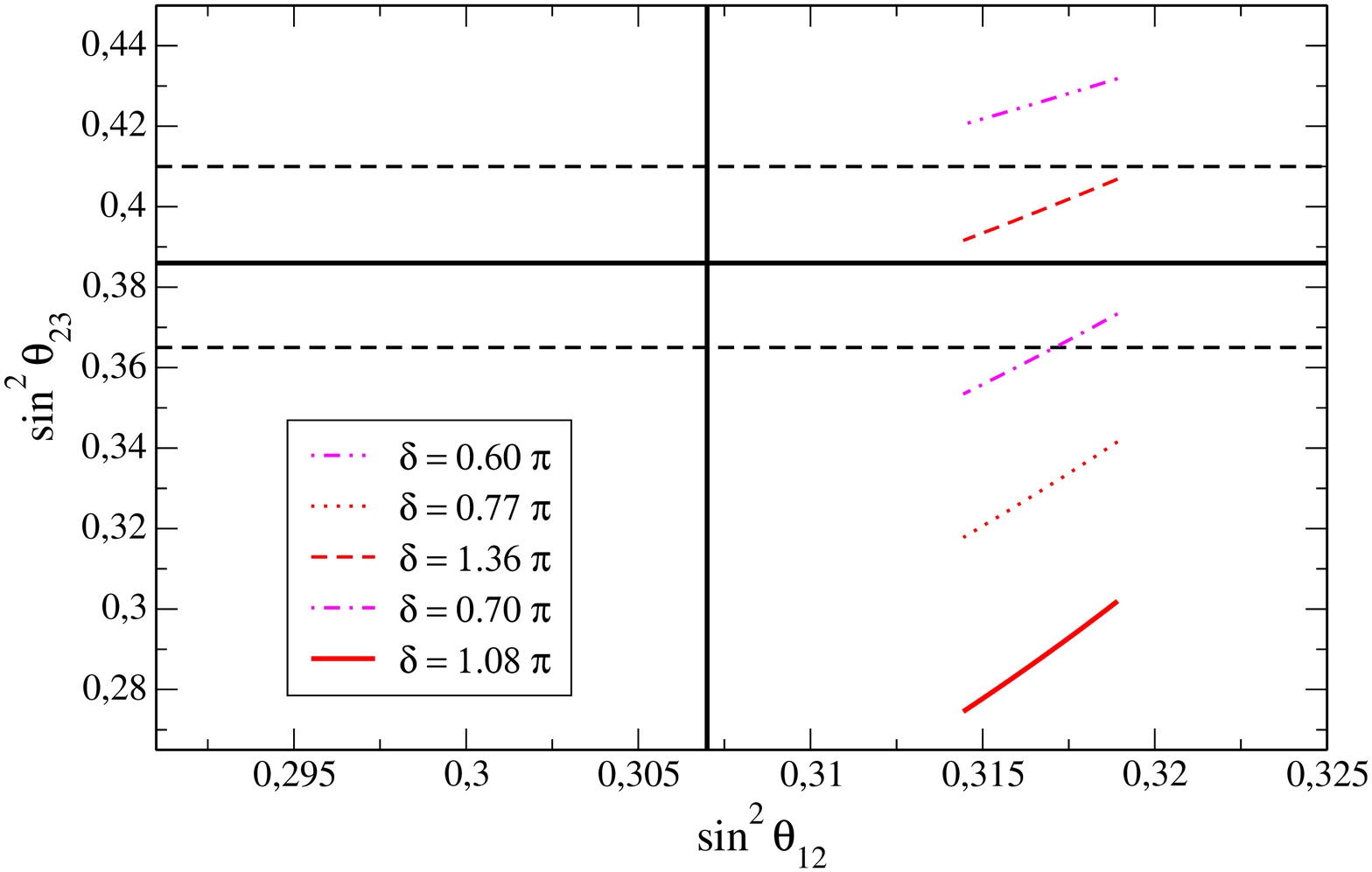}

\caption{Correlation of observables for our mixing scheme: the upper
plot shows $\sin^2 \theta_{12}$ vs.~$|U_{e3}|$, the middle plot
$\sin^2 \theta_{23}$ vs.~$\delta$ and the lower plot $\sin^2
\theta_{23}$ vs.~$\sin^2 \theta_{12}$  for different values of
$\delta$. Except for the lower plot, where the displayed range of
$\theta_{23}$ is larger, the plots cover the allowed $1\sigma$ ranges
of the parameters in case of a normal mass ordering
\cite{Fogli:2012ua}. The black solid lines are the best-fit points,
the black dashed line in the lower plot the $1\sigma$ range. }
\label{fig:plot}
\end{figure*}
\end{center}

\noindent
Consider the tri-bimaximal mixing matrix multiplied with a 23-rotation
from the right:
\be \label{eq:U}
U = U_{\rm TBM} \, R_{23}(\theta, \psi)  =
\left(
\bad
\sqrt{\frac 23} & \sqrt{\frac 13} & 0 \\
-\sqrt{\frac 16} &  \sqrt{\frac 13} &  -\sqrt{\frac 12} \\
-\sqrt{\frac 16} &  \sqrt{\frac 13} &  \sqrt{\frac 12}
\ea
\right)
\left(
\bad
1 & 0 & 0 \\
0 & \cos \theta & e^{-i \psi} \sin \theta \\
0 & -e^{i \psi} \sin \theta & \cos \theta \ea \right) . \ee The
observables are in this case \bea \D \label{eq:obs0} |U_{e3}|^2 =
\frac 13 \, \sin^2 \theta ~,~~\sin^2 \theta_{23} = \frac 12 -
\frac{\sqrt{\frac 32} \, \sin 2\theta \, \cos \psi }{3 - \sin^2
\theta}~,~~\\ \D \sin^2 \theta_{12} = 1 - \frac{2}{3 - \sin^2
\theta}~,~~ J_{\rm CP} = -\frac{1}{6\sqrt{6}} \, \sin 2 \theta \,
\sin \psi\,, \eea where the Jarlskog invariant is defined as
$J_{CP}={\rm Im} \left( U_{e1}U_{\mu 2}U^*_{e2}U^*_{\mu1}
\right)=s_{12}c_{12}s_{23}c_{23}c^2_{13}s_{13} \sin\delta$. The main
feature of the above mixing matrix is that the first column of $U$
keeps the same form as for TBM: \be \label{eq:TMv2} \left( \ba
|U_{e1}|^2 \\
|U_{\mu 1}|^2 \\
|U_{\tau 1}|^2 \ea \right) = \left( \ba 2/3 \\ 1/6 \\ 1/6 \ea
\right)  \ee From these relations, as well as from
Eq.~(\ref{eq:obs0}), the observable mixing parameters and their
correlation can be obtained. In particular, a consequence of
Eq.~(\ref{eq:TMv2}) is that $\sin^2 \theta_{12} \le \frac 13$.
Indeed, from $|U_{e1}|^2 = \frac 23$ one finds \be \sin^2
\theta_{12} = \frac 13 \, \frac{1 - 3 \, |U_{e3}|^2} {1 -
|U_{e3}|^2} \simeq \frac 13 \, \left( 1 - 2 \, |U_{e3}|^2 \right) .
\ee The desired feature b), $\sin^2 \theta_{12} = \frac 13 - {\cal
O}(|U_{e3}|^2)$, is reproduced. In Fig.~\ref{fig:plot} we show the
correlation between $|U_{e3}|$ and $\sin^2 \theta_{12}$, using the
$1\sigma$ range of the global fit results \cite{Fogli:2012ua}. For
simplicity, we use the results for the normal ordering, the
difference to the results for the inverted ordering is
insignificant. One finds that $\sin^2 \theta_{12} $ lies between
0.315 and 0.318.

The second independent condition in Eq.~(\ref{eq:TMv2}) involving
$|U_{\mu 1}|^2 = 1/6$ gives
\be \cos \delta \, \tan 2 \theta_{23} =
 - \frac{1 - 5 \, |U_{e3}|^2}{2\sqrt{2} \, |U_{e3}| \,
\sqrt{1 - 3 \, |U_{e3}|^2}} \\
\simeq \frac{-1}{2\sqrt{2} \, |U_{e3}|} \left(1 - \frac{7}{2} \,
|U_{e3}|^2 \right) . \ee This relation can be written as \be \sin^2
\theta_{23} \simeq \frac{1}{2} + \frac{1}{\sqrt{2}}  \, |U_{e3}|
\left(1 + \frac{1}{4} \, |U_{e3}|^2 \right)  \cos \delta \, . \ee
Hence, if $\cos \delta <0$ and $|\cos \delta| = {\cal O}(1)$, the
desired features c) and d) are reproduced: $\sin^2 \theta_{23} =
\frac 12 - {\cal O}(|U_{e3}|)$. We plot in Fig.~\ref{fig:plot} the
correlation between $\sin^2 \theta_{23}$ and $\delta$. To reproduce
the $1\sigma$ range of $\theta_{23}$, $\delta$ should lie between
$1.30\pi$ and $1.38\pi$. It may be useful to express the Dirac CP
phase in terms of the parameters $\theta$ and $\psi$ as
\begin{eqnarray}
\sin\delta=-\frac{\sqrt{2} (5+\cos 2 \theta ) \sin \psi
}{\sqrt{39+20 \cos 2 \theta +13 \cos 4 \theta -24 \cos 2 \psi \sin^2
2 \theta }} \; .
\end{eqnarray}
We also display in Fig.~\ref{fig:plot} the correlation between
$\sin^2 \theta_{23}$ and  $\sin^2 \theta_{12}$, when $|U_{e3}|$ is
varied in its $1\sigma$ range.

For further illustration on how nicely the mixing scheme describes
current data, we perform a somewhat naive statistical fit.
To this end, we take the two parameters $\theta$
and $\psi$ as independent, and compare the ${\rm TM}_1$
predictions to the experimental data with a $\chi^2$ function
\begin{eqnarray}
\chi^2 =\sum_{i} \frac{(\rho_i-\rho^0_i)^2}{\sigma^2_i} \, ,
\end{eqnarray}
where $\rho^0_i$ represents the data of the $i$-th experimental
observable, $\sigma_i$ the corresponding 1$\sigma$ absolute error,
and $\rho_i$ the prediction of the model. The experimental values of
the neutrino mixing angles are taken from Table I, which as
mentioned above include the data presented at Neutrino 2012. Note
that the global-fit data slightly differ for normal and inverted
neutrino mass orderings.

\begin{center}
\begin{figure*}[t]
\includegraphics[width=8cm,height=6.950cm,angle=0]{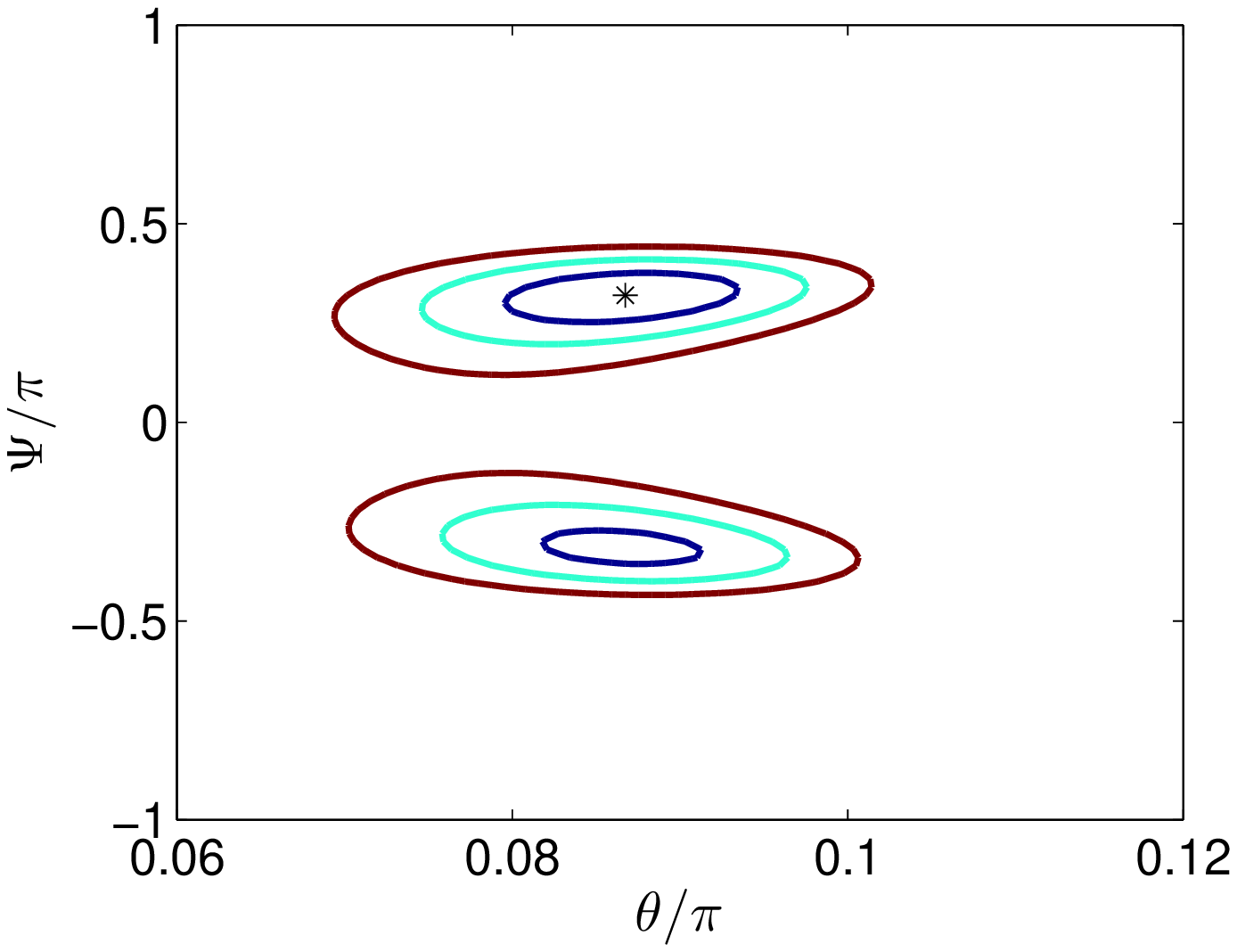}
\includegraphics[width=8cm,height=6.950cm,angle=0]{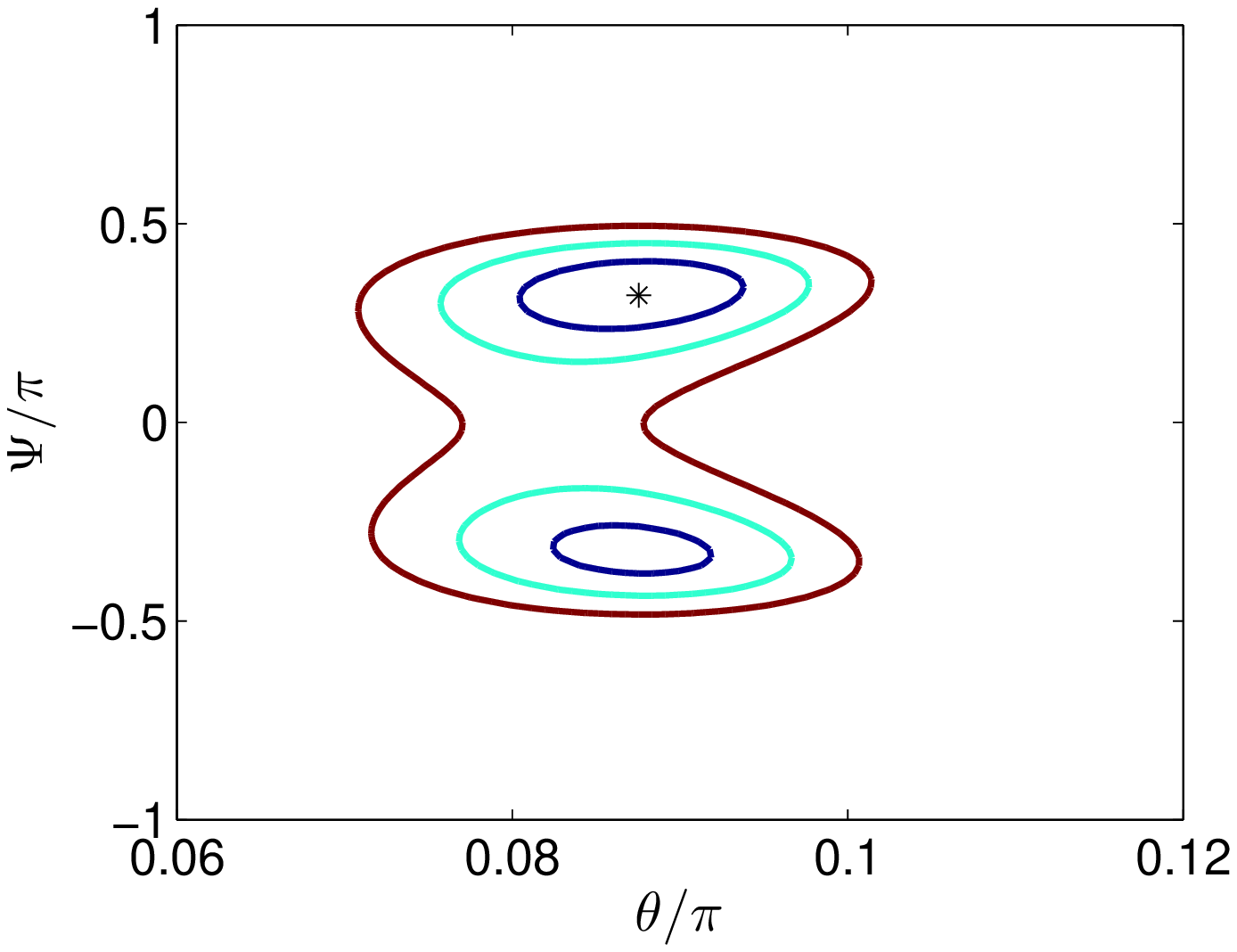}
\caption{\label{fig:psi-theta} The allowed region of the ${\rm
TM}_1$ parameters $\theta$ and $\psi$ at 1$\sigma$, 2$\sigma$, and
3$\sigma$ C.L.~for the normal mass ordering (left) and inverted mass
ordering (right). The black asterisks denote the best-fit values.
\label{fig:plot2}}
\end{figure*}
\end{center}

In Fig.~\ref{fig:psi-theta}, we present the allowed regions of
$\theta$ and $\psi$ at 1$\sigma$, 2$\sigma$, and 3$\sigma$ C.L.,
defined as contours in $\Delta \chi^2$ for two degrees of freedom
with respect to the $\chi^2$ minimum ($\chi^2_{\rm min} \simeq 1.1$
for NH and $\chi^2_{\rm min} \simeq 0.9$ for IH). Note that these
$\chi^2$ minima show that the scheme describes the data excellently.
Since $|U_{e3}|$ is firmly connected to $\theta$ in ${\rm TM}_1$, it
is restricted to a narrow range between $0.07\pi$ and $0.1\pi$, and
the best-fit value of $\theta$ lies close to $\pi/12$. The
constraints on $\psi$ are not as strong as for $\theta$ due to the
less precise determination of $\delta$ and $\theta_{23}$. The
best-fit values of $\theta$ and $\psi$ are ($0.087\pi$, $0.32\pi$)
for the NH case, and ($0.087\pi$, $0.32\pi$) for the NH case.

\begin{center}
\begin{figure*}[t]
\includegraphics[width=8cm,height=6.950cm,angle=0]{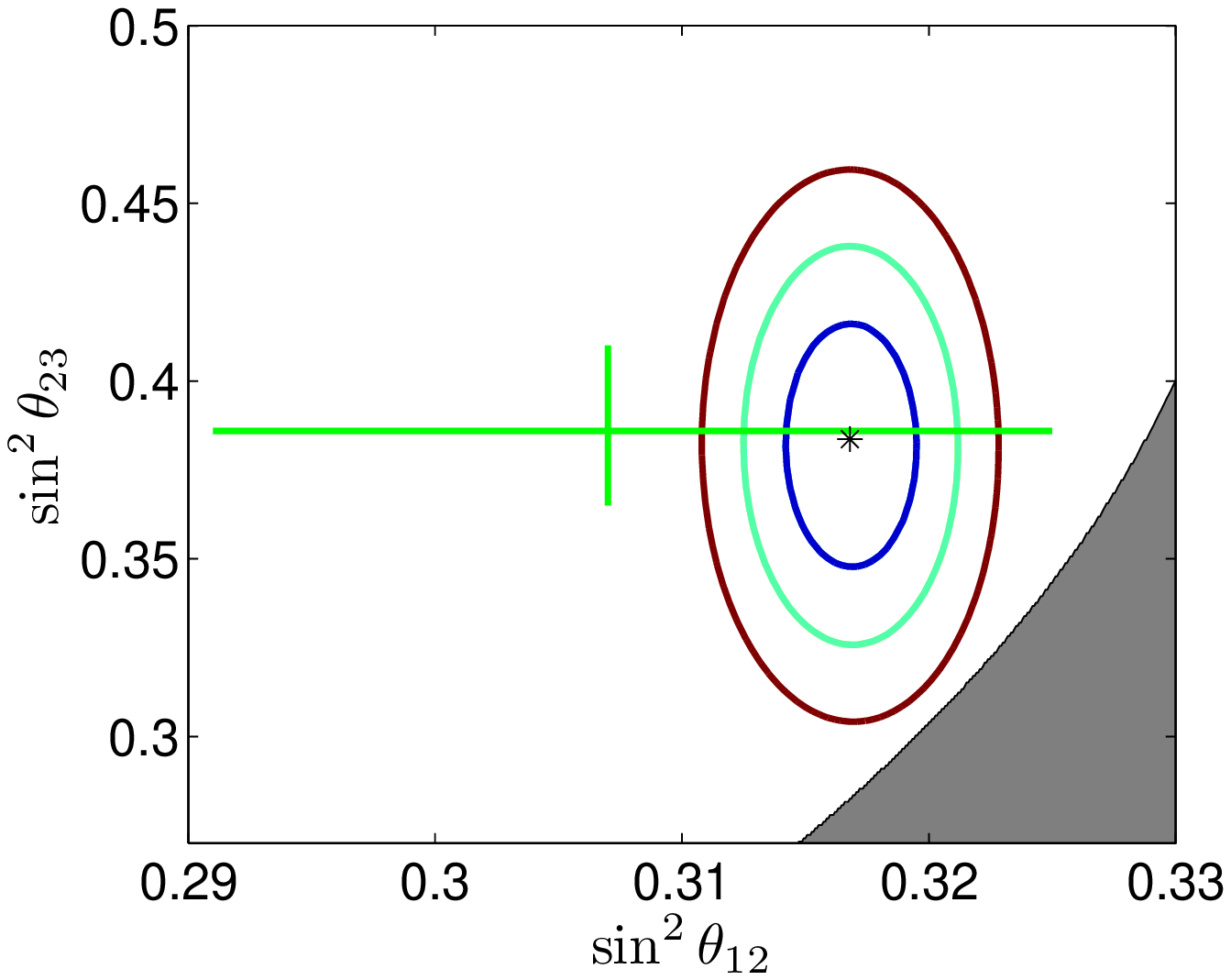}
\includegraphics[width=8cm,height=6.950cm,angle=0]{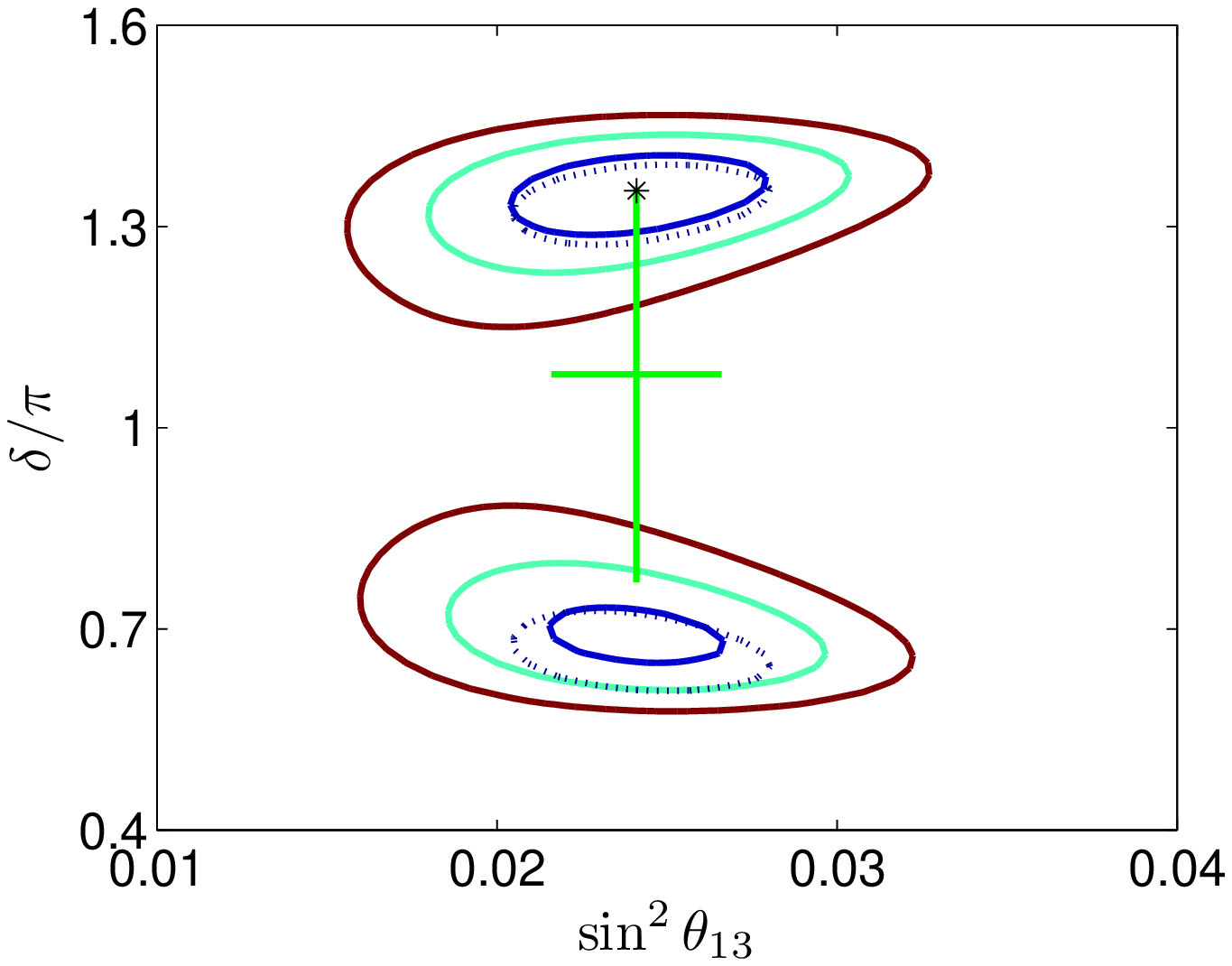}
\includegraphics[width=8cm,height=6.950cm,angle=0]{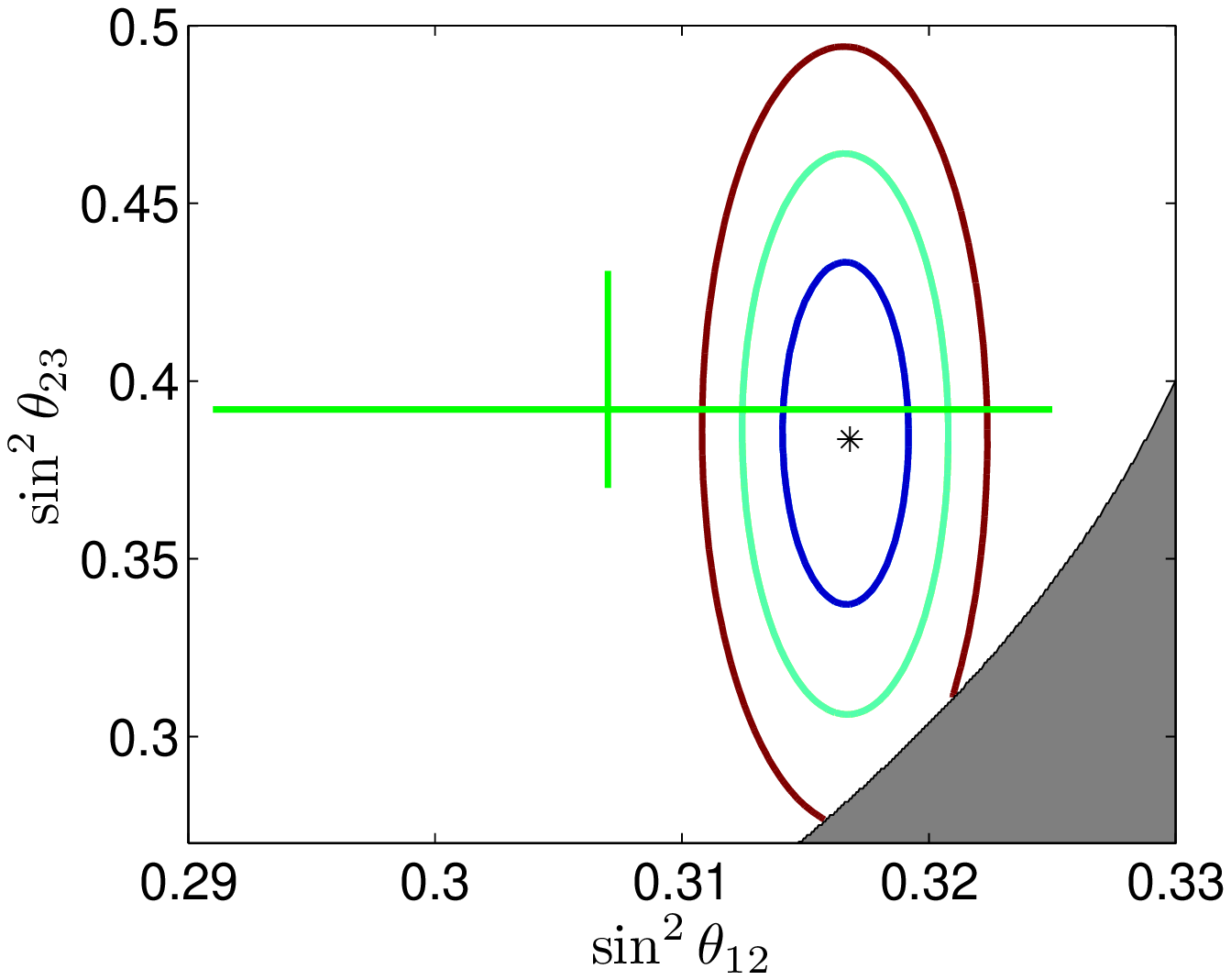}
\includegraphics[width=8cm,height=6.950cm,angle=0]{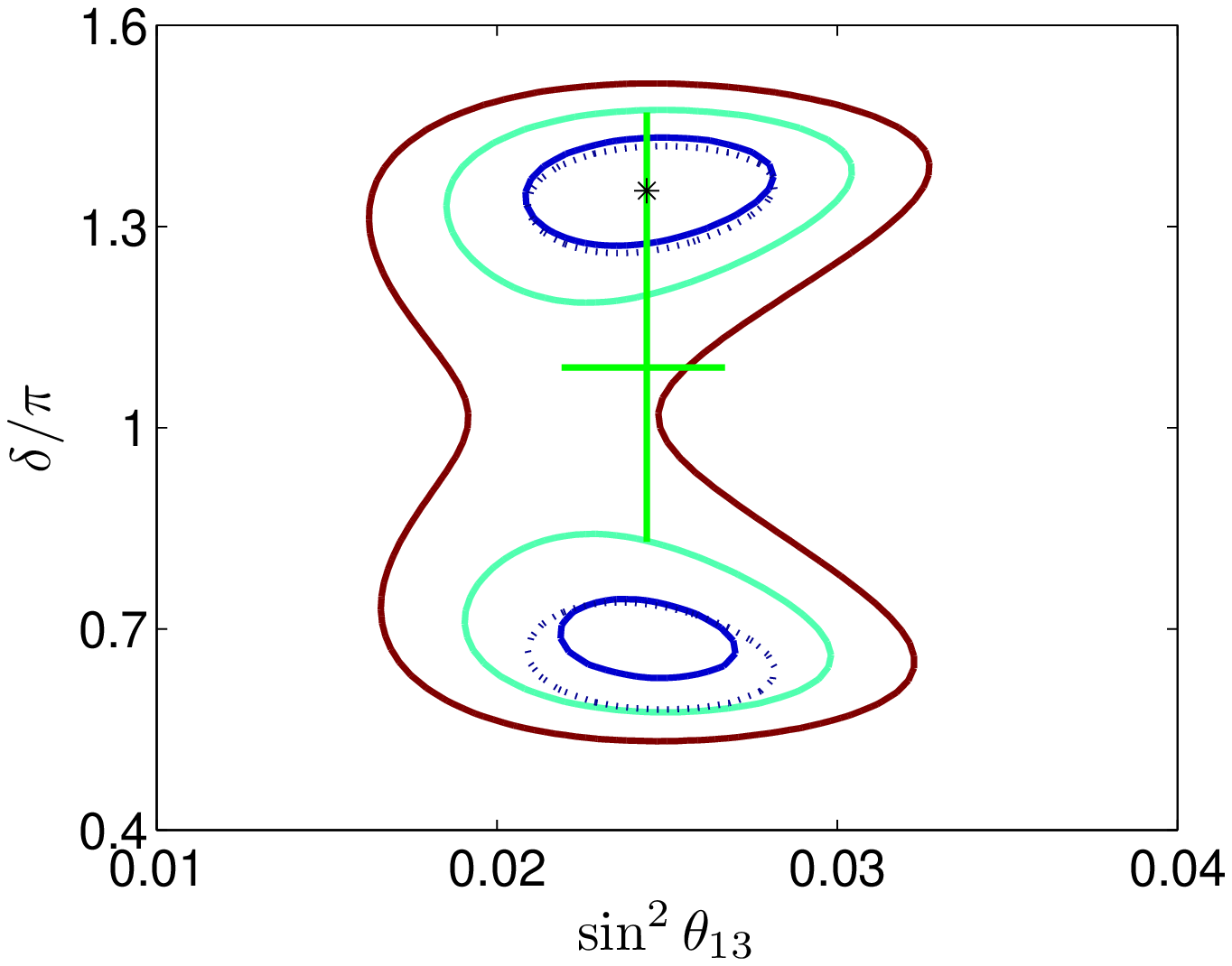}
\caption{\label{fig:parameters} The allowed region of the physical
observables at 1$\sigma$, 2$\sigma$, and 3$\sigma$ C.L.~for the
normal mass ordering (upper panel) and inverted mass ordering (lower
panel). The shaded areas are not permitted by the model. For
comparison, the allowed 1$\sigma$ ranges \cite{Fogli:2012ua} of the
parameters are also indicated by using green vertical and horizontal
bars. The dotted contours are the 1$\sigma$ ranges without
considering the experimental data on $\delta$. \label{fig:plot3}}
\end{figure*}
\end{center}

We further show in Fig.~\ref{fig:parameters} the predictions for the
mixing angles and the Dirac CP phase $\delta$ for the NH case (upper
panel) and the IH case (lower panel). Again, the data are very well
reproduced.
The best-fit values of $\sin^2\theta_{12}$ and $\sin^2\theta_{23}$
are found to be ($0.317$, $0.382$) for the NH case, and ($0.317$,
$0.385$) for the IH case. The right column shows that the ${\rm
TM}_1$ prediction on $\theta_{13}$ is in good agreement with
experiments, and that $\delta$ tends to lie at the upper end of its
allowed range. We find the best-fit values of $\sin^2\theta_{13}$
and $\delta$ to be ($0.0242$, $1.34\pi$) for NH and ($0.0245$,
$1.35\pi$) for IH. For comparison, we also show in dotted contours
the 1$\sigma$ parameter ranges without considering the experimental
data on $\delta$. In such a case, the allowed parameter spaces are
symmetric with respect to $\delta =\pi$, and $\delta$ is confined to
be either around $\delta =1.3\pi$ or $\delta =0.7\pi$, implying the
predictive power of the scenario. All in all, three mixing angles
together with one CP phase are all compatible with experimental data
within 1$\sigma$ C.L., though there are only two free parameters in
${\rm TM}_1$.

\section{\label{sec:theo}Theory}

\noindent One may ask a question arising from the previous analysis:
what is the underlying symmetry behind ${\rm TM}_1$? In this section
we wish to give some comments on the possible theoretical
background and  in particular we will show two examples in the
framework of $A_4$ and $S_4$ flavor symmetries.

\subsection{Symmetry behind the ${\rm TM}_1$}
\noindent
The original trimaximal mixing, defined by $\left(|U_{e2}|^2, |U_{\mu
2}|^2, |U_{\tau 3}|^2\right)^T = \left(1/3 , 1/3 , 1/3 \right)^T$,
has been discussed in the framework of flavor symmetry models
\cite{He:2006qd,Grimus:2008tt,Grimus:2008nf,Lin:2009bw,Grimus:2009xw,Lam:2006wm,He:2011gb,Toorop:2011jn,King:2011zj,deAdelhartToorop:2011re,Cooper:2012wf}.
Models that can lead to the mixing scheme dealt with in this paper
are, to the best of our knowledge, discussed only in
\cite{Lam:2006wm,Antusch:2011ic}.
The mass matrix that reproduces our mixing scheme is:
\be
\label{eq:TM1matrix}
\ba
  m_\nu =
U^\ast \, m^{\rm diag}_\nu \,
      U^\dagger
    = \left( \begin{array}{ccc} A & B + C & B - C\\
    \cdot & \frac{1}{2} \, (A + B + D + 4 \,C)
& \frac{1}{2} \, (A + B - D)\\
     \cdot & \cdot & \frac{1}{2} \, (A + B + D - 4 \, C)\\
      \end{array}\right) \\
= m_\nu^{\rm TBM} + \left( \bad
0 & C & -C \\
\cdot & 2 \, C & 0 \\
\cdot & \cdot & -2 \, C
\ea \right),
\ea
\ee
where we identify (with $c = \cos \theta$ and $s = \sin \theta$)
\[
\label{eq:TM1elem}
\begin{array}{rl}
  A = \frac 13 \left( 2 \, m_1 + m_2 \, c^2_\theta \, e^{-2i\alpha}
        + m_3 \, s_\theta^2 \, e^{2i(\psi - \beta)}\right)\, ,\qquad&
  C = \frac{1}{\sqrt{6}} \left(m_2 e^{-i(\psi+2\alpha)}- m_3 \,
        e^{i(\psi-2\beta)} \right) s_\theta \, c_\theta\,,\\
  B = \frac 13  \left( -m_1 + m_2 \, c^2_\theta \, e^{-2i\alpha} +
        m_3 \, s^2_\theta\, e^{2i(\psi - \beta)} \right)\,,\qquad&
  D = m_2 \, e^{-2i(\psi+\alpha)} \, s^2_\theta
+ m_3 \, c_\theta^2 \,
        e^{-2i\beta}\, .\\
\end{array}
\]
Here we have used the form of the PMNS matrix in Eq.~(\ref{eq:U}),
including for completeness the two Majorana phases. We see that the
$\mu$--$\tau$ symmetry is broken by the extra terms involving $C$.
In the limit of $C = 0$ we would have tri-bimaximal mixing. Namely,
\begin{eqnarray}
\left. m_\nu \right|_{C=0}  = m^{\rm TBM}_\nu = U_{\rm TBM} \, {\rm
diag} \left(A-B,A+2B,D\right) U^T_{\rm TBM} \; .
\end{eqnarray}
Note that $C$ is proportional to $\sin \theta$, and thus expected to
be somewhat smaller than $A,B,D$.

The eigenvalue $A-B$ has an eigenvector
$\frac{1}{\sqrt{6}}(2,-1,-1)^T$, corresponding to the invariant
column in ${\rm TM}_1$. This symmetry of the neutrino mass matrix
can be characterized by a unitary matrix $G_1$, satisfying the
relation $m_\nu \rightarrow  G^T_1 m_\nu G_1$, in which $G_1$
defines a $Z_2$ group, i.e.~$G^2_1=1$, and it is given by \be \ba
G_1 = \frac 13 \left( \bad
1 & -2 & -2 \\
-2 & -2 & 1\\
-2 & 1 & -2 \ea \right) . \ea \ee The third column of the mixing
matrix allows us to define another $Z_2$ symmetry, under which the
mass matrix is invariant,
\[
\ba G_2 = \frac 13 \left( \bad
2 + c_{2 \theta} & s_\theta (\sqrt{6} e^{i \psi} c_\theta - 2
s_\theta ) & -s_\theta (\sqrt{6} e^{i \psi} c_\theta +
2 s_\theta ) \\
s_\theta (\sqrt{6} e^{-i \psi} \, c_\theta - 2 s_\theta ) & s_\theta
(2\sqrt{6} c_\theta \cos \psi + s_\theta ) & \frac 12 (1 + 5 c_{2
\theta} - 2 i \sqrt{6} \sin \psi s_{2 \theta})
\\
-s_\theta (\sqrt{6} e^{-i \psi}  c_\theta + 2 s_\theta ) & \frac 12
(1 + 5 c_{2 \theta} + 2 i \sqrt{6} \sin \psi s_{2 \theta}) &
-s_\theta (2\sqrt{6} c_\theta \cos \psi - s_\theta ) \ea \right) .
\ea
\]
Therefore, the neutrino mass matrix is invariant under the $Z_2
\times Z_2$ transformation generated by $G_1\times G_2$. Note that
we are working in the basis in which the charged lepton mass matrix
is diagonal and non-degenerate. Therefore, any unitary matrix $F$
that commutes with the charged lepton mass matrix must be diagonal
with unit moduli in all its entries. For the three generation case,
there are three distinct $F$: ${\rm diag}(1,\omega,\omega^2)$, ${\rm
diag}(\omega,1,\omega^2)$ and ${\rm diag}(\omega^2,1,\omega)$ with
$\omega=\exp(2\pi i/3)$. This allows one to construct a flavor group
$\cal G$ generated by $F$, $G_1$ and $G_2$, i.e.~${\cal
G}=\left\{F,G_1,G_2\right\}$. The generator $G_1$ is well motivated
in several flavor symmetry groups, including $A_4$, $S_4$ and other
higher order discrete groups (see Ref.~\cite{Lam:2006wm} for a
detailed discussion, where in fact $S_4$ was proposed for the scheme
under consideration). In contrast, $G_2$ cannot be simply embedded
into small discrete groups due to the rotation angle $\theta$. For
$\theta = 0$ the matrix $G_2$ becomes the generator of $\mu$--$\tau$
symmetry, \be G_2 (\theta=0) = \left( \bad
1 & 0 & 0 \\
0 & 0 & 1 \\
0 & 1 & 0 \ea \right).
\ee
For $\theta=\pi/12$ (which is close to the best-fit point) and $\phi=\pi$, one has
\begin{eqnarray}
G_2 (\theta=\pi/12) = \begin{pmatrix}
 \frac{1}{6} \left(4+\sqrt{3}\right) & \frac{1}{12} \left(-4+2 \sqrt{3}-\sqrt{6}\right) & \frac{1}{12} \left(-4+2 \sqrt{3}+\sqrt{6}\right)
 \cr
 \frac{1}{12} \left(-4+2 \sqrt{3}-\sqrt{6}\right) & \frac{1}{12} \left(2-\sqrt{3}-2 \sqrt{6}\right) & \frac{1}{12} \left(2+5 \sqrt{3}\right)
 \cr
 \frac{1}{12} \left(-4+2 \sqrt{3}+\sqrt{6}\right) & \frac{1}{12} \left(2+5 \sqrt{3}\right) & \frac{1}{6}-\frac{1}{4 \sqrt{3}}+\frac{1}{\sqrt{6}}
\end{pmatrix}  ,
\end{eqnarray}
which actually gives $|U_{e3}|^2\simeq 0.025$, $\sin^2\theta_{12}
\simeq 0.318$ and $\sin^2\theta_{23} \simeq 0.709$. Both
$\theta_{13}$ and $\theta_{12}$ are well within the current
$1\sigma$ ranges, while $\theta_{23}$ deviates from its best-fit
value at more than $2\sigma$ C.L., which might be improved once an
explicit model with perturbations is constructed.

\subsection{Realization in flavor symmetry models}
\noindent As we mentioned above, the ${\rm TM}_1$ mixing scheme can
be viewed as a modification to the TBM mixing pattern by multiplying
a $23$-rotation from the right to $U_{\rm TBM}$. In this
sense, the rotation matrix $R_{23}(\theta,\psi)$ could also be
viewed as a perturbation to the exact TBM mixing
pattern. Alternatively, and this is what happens in the two short
examples we are about to give, one can note that the mass matrix for
${\rm TM}_1$ is the TBM mass matrix plus a simple additional term, see
Eq.~(\ref{eq:TM1matrix}). It is therefore possible that we modify a
successful model leading to TBM, and add additional flavons and
particles to it which give precisely the required form for TM$_1$.

In the original Altarelli--Feruglio model
\cite{Altarelli:2005yp,Altarelli:2005yx}, the left-handed lepton
doublets are assigned to a three dimensional representation $\bf 3$
under the tetrahedral group $A_4$, whereas the right-handed lepton
fields transform as $\bf 1$, $\bf 1''$ and $\bf 1'$, respectively.
In addition, three sets of flavon fields $\varphi_T$, $\varphi_S$
and $\xi$, transforming as $\bf 3$, $\bf 3$ and $\bf 1$ under $A_4$,
are also introduced together with the vacuum expectation
values\footnote{Note that we assume the solution of the vacuum
alignment could be achieved. In fact, several methods (e.g.~by using
driving fields) have been proposed to explain the different
alignments, and these will not be discussed here.} (VEVs): $\langle
\varphi_T \rangle = (v_T,0,0)$, $\langle \varphi_S \rangle =
(v_S,v_S,v_S)$ and $\langle \xi \rangle = u$, respectively. At
leading order, the $A_4$ invariant Lagrangian contains terms like
\bea \D
 {\cal L}  =  y_e \frac{1
}{\Lambda} e^c \left(  \varphi_T \ell \right) h_d + y_\mu \frac{1
}{\Lambda} \mu^c \left(  \varphi_T \ell \right)'h_d + y_\tau\frac{
1}{\Lambda} \tau^c \left(  \varphi_T \ell \right)''  h_d  \\
 + \frac{1}{\Lambda^2} x_a \xi \left( \ell h_u \ell h_u \right)+
\frac{1}{\Lambda^2} x_b \varphi_S \left( \ell h_u \ell h_u \right)+
{\rm h.c.}, \eea where $y_\alpha$ and $x_i$ denote the corresponding
Yukawa couplings, $\Lambda$ is the cut-off scale of the theory, and
two Higgs doublets $h_u$ and $h_d$ with VEVs $v_u$ and $v_d$ are
assumed to be invariant under $A_4$. Note that there is also an
additional $Z_3$ symmetry in the model, which decouples the charged
lepton and neutrino sectors
\cite{Altarelli:2005yp,Altarelli:2005yx}. By inserting the VEVs of
the flavon fields, one finds that the charged lepton mass matrix is
diagonal at leading order, i.e.
\begin{eqnarray}
m^{(0)}_\ell & = & \frac{v_d v_T}{\Lambda} {\rm diag }\left(  y_e ,
y_\mu , y_\tau \right)  ,
\end{eqnarray}
whereas the neutrino mass matrix is given by
\begin{eqnarray}
m^{(0)}_\nu & =& \frac{v^2_u}{\Lambda} \begin{pmatrix} a +
\frac{2b}{3} & -\frac{b}{3} & -\frac{b}{3} \cr -\frac{b}{3} &
\frac{2b}{3} &  a -\frac{b}{3} \cr -\frac{b}{3} & a -\frac{b}{3} &
\frac{2b}{3}
\end{pmatrix}  ,
\end{eqnarray}
with $a=2x_a \frac{u}{\Lambda}$ and $b=2x_b \frac{v_S}{\Lambda}$. The
leading order mass matrix
$m^{(0)}_\nu$ is then diagonalized by $U_{\rm TBM}$ as
\begin{eqnarray}
m^{(0)}_\nu = \frac{v^2_u}{\Lambda}  U_{\rm TBM} {\rm diag} \left(
a+b , a , -a + b \right)  U^T_{\rm TBM} \; .
\end{eqnarray}
Note that $m^{(0)}_\nu$ is consistent with the TBM mass matrix form
$m^{\rm TBM}_\nu$ defined in Eq.~\eqref{eq:TM1matrix} since the
identification $A=a+\frac{2}{3}b$, $B=-\frac{1}{3}b$ and $D=b-a$ can
be made.

In order to modify the TBM mixing pattern, we introduce another
flavon field $\phi$, which transforms as $\bf 3$ and couples to the
lepton doublets via $\frac{1}{\Lambda^2} x_c  \phi \left( \ell h_u
\ell h_u \right)$. Similar to $\xi$ and $\varphi_S$, the unwanted
couplings between right-handed charged leptons and $\phi$ are
forbidden by the additional $Z_3$ symmetry. Different from the
flavons $\varphi_T$ and $\varphi_S$, the flavon field $\phi$ is
assumed to develop a VEV along the directions $\langle \phi \rangle
= (0,-v_\phi,v_\phi)$. This vacuum alignment follows the
orthogonality conditions $\langle \phi \rangle \cdot \langle
\varphi_T \rangle$ and $\langle \phi \rangle \cdot \langle \varphi_S
\rangle$, where the ``$\cdot$'' denotes the usual scalar product of
3-vectors. It has been shown in Ref.~\cite{Antusch:2011ic} that such
an orthogonality condition can be realized within supersymmetry with
``Lagrange multiplier'' superfields, which are singlets under the
flavor symmetry but couple to the flavon fields in the
superpotential. The $F$-term conditions, which are equivalent to the
orthogonality conditions, could then yield the desired vacuum
alignments.

The residual symmetry in the neutrino sector is now broken by $\langle \varphi_S \rangle =
(v_S,v_S,v_S)$ down to a $Z_2$ symmetry $G_1$. The additional term for
$m_\nu$
from the extra flavon, after acquiring its VEV alignment $\langle \phi
\rangle = (0,-v_\phi,v_\phi)$, gives
\begin{eqnarray}
m^{(1)}_\nu =  \frac{v^2_u}{\Lambda} \begin{pmatrix} 0 & \frac{c}{2}
& -\frac{c}{2} \cr \frac{c}{2} & c & 0 \cr -\frac{c}{2} & 0 & -c
\end{pmatrix} ,
\end{eqnarray}
where $c=2x_c \frac{v_\phi}{\Lambda}$. Writing $C=\frac{1}{2}c$, the
${\rm TM}_1$ matrix structure given in Eq.~\eqref{eq:TM1matrix} is
then reproduced.

The above analysis could also be applied to other groups containing
$A_4$ as a subgroup. In the $S_4$ model explored in
Ref.~\cite{Bazzocchi:2009pv}, the lepton doublets are assigned to a
three dimensional representation ${\bf 3}_1$ under $S_4$ as well as
two flavons $\psi \sim {\bf 2}$ and $\Delta \sim {\bf 3}_1$. The
vacuum alignments are taken to be $\langle \psi \rangle =
\left(v_\psi , v_\psi \right)$ and $\langle \Delta \rangle =
\left(v_\Delta , v_\Delta, v_\Delta \right)$. The neutrino mass
matrix then reads
\begin{eqnarray}
m^{(0)}_\nu =  \frac{v^2_u}{\Lambda} \begin{pmatrix} 2f & d-f & d-f
\cr d-f & d+2f & -f \cr d-f & -f & d+2f
\end{pmatrix} ,
\end{eqnarray}
where $d= 2x_d v_\psi/\Lambda$ and $f= 2x_f v_\Delta /\Lambda$ stem
from the Yukawa coupling terms. This is equivalent to TBM, which can
be seen by writing $A=2f$, $B=d-f$ and $D=d+3f$. Hence,
$m^{(0)}_\nu$ is diagonalized as
\begin{eqnarray}
m^{(0)}_\nu = \frac{v^2_u}{\Lambda}  U_{\rm TBM} {\rm diag} \left(
-d+3f, 2d , d + 3f \right)  U^T_{\rm TBM} \, .
\end{eqnarray}
Similar to the $A_4$ model, here we introduce a new flavon field
$\zeta$, which transforms as ${\bf 3}_1$ but possesses a special
vacuum structure, $\langle \zeta \rangle \sim \left(
0,-v_\zeta,v_\zeta\right) $, again possible to achieve with an
orthogonality condition. The new flavon field leads to the
following contribution to the neutrino mass term
\begin{eqnarray}
m^{(1)}_\nu =  \frac{v^2_u}{\Lambda} \begin{pmatrix} 0 & s& -s \cr s
& 2s & 0 \cr -s & 0 & -2s
\end{pmatrix}  ,
\end{eqnarray}
where $s=y v_\zeta/\Lambda$ with $y$ being the Yukawa coupling
between $\zeta$ and the lepton doublets. Again, when $m^{(1)}_\nu$
is added to $m^{(0)}_\nu$, the ${\rm TM}_1$ mass matrix is
reproduced.

In these two examples, the VEVs of the new flavon
fields should in principle be smaller than the other flavon VEVs, the
reason being that $C$ is proportional to $\sin \theta$ and expected to
be suppressed with respect to the other entries in
Eq.~\eqref{eq:TM1matrix}.
This could be achieved if the scale of new flavon fields is roughly
one order of magnitude smaller than the others.

\section{\label{sec:concl}Summary}
\noindent In this work we have studied an attractive neutrino mixing
scheme ${\rm TM}_1$, in which the first column of the PMNS matrix
has the same form as for tri-bimaximal mixing. The PMNS matrix can
be described by using only two parameters: one rotation angle
$\theta$ and one CP phase $\psi$. The physical observables, i.e.~the
three mixing angles and the Dirac phase, are therefore correlated
via the two parameters, leaving us with rather definite
phenomenology. While this was studied before, we noted here that the
features that apparently emerge from global fits can be excellently
described by this mixing scheme. Namely, if $|U_{e3}|$ is non-zero,
solar neutrino mixing is governed by $\sin^2 \theta_{12} = \frac 13
- {\cal}(|U_{e3}|^2)$, i.e.~slightly less than $\frac13$. If in
addition the CP phase $\delta$ is such that $\cos \delta$ is
negative, and located around $\pi$, then atmospheric neutrino mixing
is governed by $\sin^2 \theta_{23} = \frac 13 - {\cal}(|U_{e3}|)$,
i.e.~significantly less than $\frac 12$. These features, small
negative deviations from $\frac13$, large negative deviations from
$\frac 12$ and $\delta$ around $\pi$, are (within $1\sigma$) the
outcome of the global fit that we referred to. In fact, the sizable
deviation from maximal mixing requires that the CP phase $\delta$ is
around $\pi$, which can be tested in upcoming long-baseline neutrino
facilities. We have also discussed potential flavor symmetries
behind the ${\rm TM}_1$ scheme, and showed in particular that rather
straightforward additions to existing models leading to
tri-bimaximal mixing, be it $A_4$ or $S_4$, can lead to the mixing
scheme.

It will be interesting to see whether these features to which global
fits seem to point survive the test of time. The mixing scheme that
we studied here seems to be very well suited to describe the current
data, and its rather simple structure adds to its attractiveness.

\vspace{0.3cm}
\begin{center}
{\bf Acknowledgments}
\end{center}
This work is supported by the ERC under the Starting Grant
MANITOP.



\begin{thebibliography}{99}







\bibitem{Abe:2011fz}
  Y.~Abe {\it et al.}  [DOUBLE-CHOOZ Collaboration],
  Phys.\ Rev.\ Lett.\  {\bf 108}, 131801 (2012)
  [arXiv:1112.6353 [hep-ex]].

\bibitem{An:2012eh}
  F.~P.~An {\it et al.}  [DAYA-BAY Collaboration],
  Phys.\ Rev.\ Lett.\  {\bf 108}, 171803 (2012)
  [arXiv:1203.1669 [hep-ex]].


\bibitem{Ahn:2012nd}
  J.~K.~Ahn {\it et al.}  [RENO Collaboration],
  Phys.\ Rev.\ Lett.\  {\bf 108}, 191802 (2012)
  [arXiv:1204.0626 [hep-ex]].


\bibitem{Machado:2011ar}
  P.~A.~N.~Machado, H.~Minakata, H.~Nunokawa and R.~Z.~Funchal,
  JHEP {\bf 1205}, 023 (2012)
  [arXiv:1111.3330 [hep-ph]].

\bibitem{Tortola:2012te}
  M.~Tortola, J.~W.~F.~Valle and D.~Vanegas,
  arXiv:1205.4018 [hep-ph].


\bibitem{Fogli:2012ua}
  G.~L.~Fogli, E.~Lisi, A.~Marrone, D.~Montanino, A.~Palazzo and A.~M.~Rotunno,
  arXiv:1205.5254 [hep-ph].



\bibitem{thomas}T.~Schwetz, talk at NuTURN 2012, see {\tt
agenda.infn.it/conferenceDisplay.py?confId=4722}


\bibitem{tbm}P.~F.~Harrison, D.~H.~Perkins and W.~G.~Scott,
  Phys.\ Lett.\ B {\bf 530}, 167 (2002);
  Phys.\ Lett.\ B {\bf 535}, 163 (2002);
Z.~Z.~Xing,
  Phys.\ Lett.\ B {\bf 533}, 85 (2002);
  X.~G.~He and A.~Zee,
  Phys.\ Lett.\ B {\bf 560}, 87 (2003).


\bibitem{Albright:2008rp}
  C.~H.~Albright and W.~Rodejohann,
  Eur.\ Phys.\ J.\ C {\bf 62}, 599 (2009)
  [arXiv:0812.0436 [hep-ph]];
see also C.~H.~Albright, A.~Dueck and W.~Rodejohann,
  Eur.\ Phys.\ J.\ C {\bf 70}, 1099 (2010)
  [arXiv:1004.2798 [hep-ph]].

\bibitem{He:2011gb}
  X.~-G.~He and A.~Zee,
  Phys.\ Rev.\ D {\bf 84}, 053004 (2011)
  [arXiv:1106.4359 [hep-ph]].



\bibitem{Bjorken:2005rm}
  J.~D.~Bjorken, P.~F.~Harrison and W.~G.~Scott,
  Phys.\ Rev.\ D {\bf 74}, 073012 (2006)
  [hep-ph/0511201].


\bibitem{He:2006qd}
  X.~-G.~He and A.~Zee,
  Phys.\ Lett.\ B {\bf 645}, 427 (2007)
  [hep-ph/0607163].

\bibitem{Lam:2006wm}
  C.~S.~Lam,
  Phys.\ Rev.\ D {\bf 74}, 113004 (2006)
  [hep-ph/0611017];
Phys.\ Lett.\ B {\bf 656}, 193 (2007)
  [arXiv:0708.3665 [hep-ph]];
  arXiv:0907.2206 [hep-ph];




\bibitem{Grimus:2008tt}
  W.~Grimus and L.~Lavoura,
  JHEP {\bf 0809}, 106 (2008)
  [arXiv:0809.0226 [hep-ph]].


\bibitem{Grimus:2008nf}
  W.~Grimus and L.~Lavoura,
  Phys.\ Lett.\ B {\bf 671}, 456 (2009)
  [arXiv:0810.4516 [hep-ph]].

\bibitem{Lin:2009bw}
  Y.~Lin,
  Nucl.\ Phys.\ B {\bf 824}, 95 (2010)
  [arXiv:0905.3534 [hep-ph]].


\bibitem{Grimus:2009xw}
  W.~Grimus, L.~Lavoura and A.~Singraber,
  Phys.\ Lett.\ B {\bf 686}, 141 (2010)
  [arXiv:0911.5120 [hep-ph]].






\bibitem{Toorop:2011jn}
  R.~d.~A.~Toorop, F.~Feruglio and C.~Hagedorn,
  Phys.\ Lett.\ B {\bf 703}, 447 (2011)
  [arXiv:1107.3486 [hep-ph]].



\bibitem{King:2011zj}
  S.~F.~King and C.~Luhn,
  JHEP {\bf 1109}, 042 (2011)
  [arXiv:1107.5332 [hep-ph]].


\bibitem{deAdelhartToorop:2011re}
  R.~de Adelhart Toorop, F.~Feruglio and C.~Hagedorn,
  Nucl.\ Phys.\ B {\bf 858}, 437 (2012)
  [arXiv:1112.1340 [hep-ph]].

\bibitem{Cooper:2012wf}
  I.~K.~Cooper, S.~F.~King and C.~Luhn,
  arXiv:1203.1324 [hep-ph].

\bibitem{Antusch:2011ic}
  S.~Antusch, S.~F.~King, C.~Luhn and M.~Spinrath,
  Nucl.\ Phys.\ B {\bf 856}, 328 (2012)
  [arXiv:1108.4278 [hep-ph]].



\bibitem{Altarelli:2005yp}
  G.~Altarelli and F.~Feruglio,
  Nucl.\ Phys.\ B {\bf 720} (2005) 64
  [hep-ph/0504165].

\bibitem{Altarelli:2005yx}
  G.~Altarelli and F.~Feruglio,
  Nucl.\ Phys.\ B {\bf 741} (2006) 215
  [hep-ph/0512103].

\bibitem{Bazzocchi:2009pv}
  F.~Bazzocchi, L.~Merlo and S.~Morisi,
  Nucl.\ Phys.\ B {\bf 816} (2009) 204
  [arXiv:0901.2086 [hep-ph]].



\end{thebibliography}
\end{document}